\documentclass[11pt]{article}
\usepackage{graphicx}
\usepackage{epstopdf}
\AppendGraphicsExtensions{.tif}

\usepackage[english]{babel}
\usepackage[ansinew]{inputenc}
\usepackage{epsfig}
\usepackage{float}
\usepackage{bm}
\usepackage{fullpage}
\usepackage{wrapfig}
\usepackage{float}
\usepackage{amsfonts}
\usepackage{amssymb}
\usepackage{colortbl}
\usepackage{gensymb}
\usepackage{amsmath}
\usepackage{natbib}
\usepackage{amssymb}

\graphicspath{{./Figures/}}

\begin{document}

\title{
Optogenetic vision restoration with high resolution
}

\author{Ferrari U.$^{1}$*, Deny S.$^{1,2}$*, Sengupta A.$^{1}$*, Caplette R.$^{1}$, Sahel, J.A.$^{1}$, Dalkara D.$^{1}$, \\ Picaud S.$^{1}$+, Duebel J.$^{1}$+, Marre O.$^{1}$+}

\maketitle

\small $^{1}$Sorbonne Universite, INSERM, CNRS, Institut de la Vision, 17 rue Moreau, F-75012 Paris, France. \\
\small $^{2}$Current affiliation: Department of Applied Physics, Stanford University. \\
*,+: equal contributions\\
\\

\abstract{ The majority of inherited retinal degenerations are due to photoreceptor cell death. In many cases ganglion cells are spared making it possible to stimulate them to restore visual function. 
Several studies \citep{Bi2006,Lin2008,Sengupta2016,Caporale2011,Berry2017} have shown that it is possible to express an optogenetic protein in ganglion cells and make them light sensitive. 
This is a promising strategy to restore vision since optical targeting may be more precise than electrical stimulation with a retinal prothesis. 
However the spatial resolution of optogenetically-reactivated retinas has not been measured with fine-grained stimulation patterns. Since the optogenetic protein is also expressed in axons, it is unclear if these neurons will only be sensitive to the stimulation of a small region covering their somas and dendrites, or if they will also respond to any stimulation overlapping with their axon, dramatically impairing spatial resolution. Here we recorded responses of mouse and macaque retinas to random checkerboard patterns following an in vivo optogenetic therapy. We show that optogenetically activated ganglion cells are each sensitive to a small region of visual space. A simple model based on this small receptive field predicted accurately their responses to complex stimuli. From this model, we simulated how the entire population of light sensitive ganglion cells would respond to letters of different sizes. We then estimated the maximal acuity expected by a patient, assuming it could make an optimal use of the information delivered by this reactivated retina. The obtained acuity is above the limit of legal blindness. This high spatial resolution is a promising result for future clinical studies.}
\\\\\\

\newpage

\section*{Results}

\subsection*{Optogenetically activated ganglion cells have localized receptive fields}

We targeted retinal ganglion cells (RGCs) of blind rd1 mice (4/5 weeks old) with an AAV2 encoding ReaChR-mCitrine (a variant of Channel Rhodospin with red-shifted sensitivity) under a pan-neuronal hSyn promoter via intravitreal injections. Details of the gene delivery and optogenetic protein expression has been detailed elsewhere \citep{Sengupta2016,Chaffiol2017a}. Retinas were harvested after 4 weeks for multi-electrode array (MEA) recordings. \\

\begin{figure}
\begin{center}
\includegraphics[width=\linewidth]{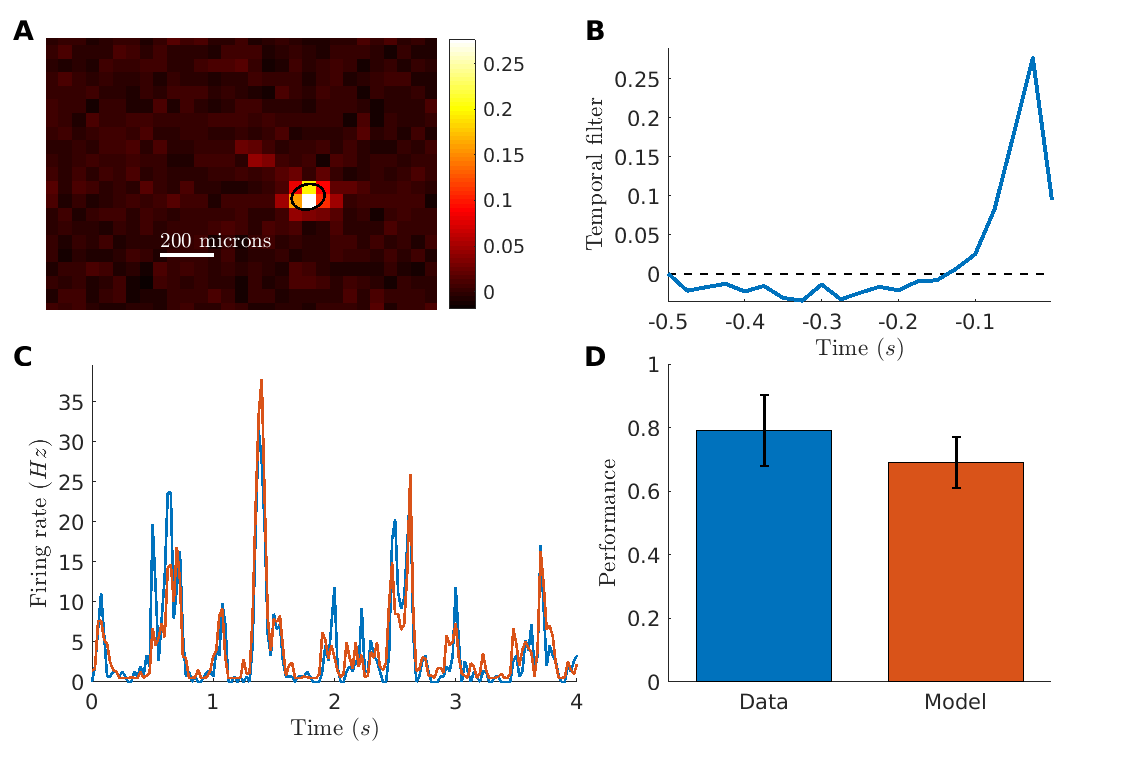}
\end{center}
\caption{\textbf{Receptive field and LN model for a ReaChR reactivated mouse retina.} (A) Spatial receptive field of one example RGC, computed from the spike-triggered average (STA) to a checkerboard stimulus. Black ellipse is the 1-std contour of a gaussian fit to the STA. Scale bar: 200 microns. (B) Temporal receptive field computed from the STA of one RGC (see methods). (C) Average firing rate (blue) from a cell in response to a repeated sequence of the checkerboard stimulus and predicted firing rate (red) from a LN model fitted to the cell response (r$=$ 0.82). (D) Prediction performance of the model against reliability of the response (see methods).  
}
\label{FigMouseRF}
\end{figure}

To estimate the size of the region whose stimulation can activate a ganglion cell (i.e. its receptive field), we displayed a random checkerboard stimulus (see methods). We estimated the Spike Triggered Average (STA) by averaging over the frames that evoked a spike (see methods). Many cells showed a well-defined receptive field (fig. \ref{FigMouseRF}A). The average diameter of receptive fields was 98.7$\pm$7 $\mu$m (SEM, n$=$30), slightly smaller than what is usually measured in normal retinas \citep{Farrow2011}. The temporal time course of the receptive field (fig. \ref{FigMouseRF}B) was consistent with a direct light-activation of ganglion cells: the latency of the peak response was around 25 ms. \\

To characterize the processing performed by the reactivated retinas, we built a classical Linear-Non-linear (LN) model \citep{Chichilnisky_2001} to predict the responses of ganglion cells to the checkerboard stimulus. The LN model is composed of a linear filter followed by a non-linearity to predict the firing rate of the cell. The linear filter was the average STA measured with the checkerboard. The non-linearity was learned on the checkerboard data using a classical maximum-likelihood estimation (see methods). 

We then tested the model on a repeated sequence of the checkerboard stimulus that was not used to learn the model. We restricted our analysis to ganglion cells that were strongly modulated by the stimulus (reliability score above $0.5$, see methods), which correspond to ganglion cells where the expression level of the optogenetic protein was high enough. 
On these cells, the LN model predicted very well cell responses to the checkerboard stimulus (fig. \ref{FigMouseRF}C for an example). For all these cells, the prediction performance was high (mean Pearson correlation  0.69 $\pm$ 0.08, n$=$24), close to the limit set by the reliability of the response (fig.  \ref{FigMouseRF}D and methods).\\

\begin{figure}
\begin{center}
\includegraphics[width=\linewidth]{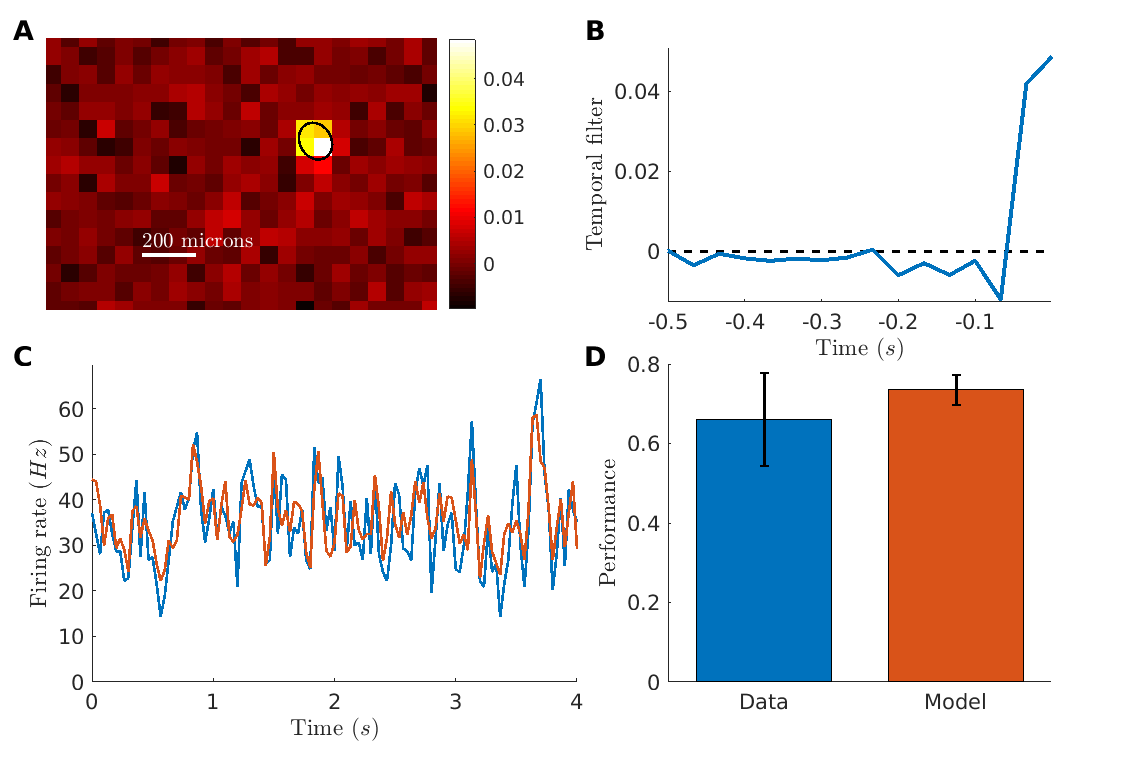}
\end{center}
\caption{ 
\textbf{Receptive field and LN model for a CatCh reactivated macaque retina.} (A) Spatial receptive field of one example RGC, computed from the spike-triggered average (STA) to a checkerboard stimulus. Black ellipse is the 1-std contour of a Gaussian fit to the STA. Scale bar: 200 microns. (B) Temporal receptive field computed from the STA of one RGC (see methods). (C) Average firing rate (blue) from a cell in response to a repeated sequence of the checkerboard stimulus and predicted firing rate (red) from a LN model fitted to the cell response (r$=$ 0.77). (D) Prediction performance of the model against reliability of the response (see methods).  
}
\label{FigMonkeyRF}
\end{figure}

We then tested if the same approach could be applied in the non-human primate retina. We used data from a macaque retina where ganglion cells were transfected with the optogenetic protein CatCh (a variant of Channel-rhodopsin) \citep{Chaffiol2017a}. Here AAV2 vectors were produced with a promoter driving the expression of the CatCh protein in ganglion cells. Macaques were injected intravitreally with AAV particles. Retinas were harvested 12 weeks after injection for MEA recordings. To suppress the endogeneous photoreceptor response, retinae were light exposed for several hours, and LAP4 was added to the bath to block ON responses due to photoreceptor (see methods). We performed a similar experiment than for the mouse, displaying a random checkerboard and measuring receptive fields. Although the activity modulation by visual stimuli was much weaker than for the mouse (spontaneous activity was high, probably an effect of the culture or the bleaching), we could still find STA with a well-defined receptive field in some ganglion cells (fig.\ref{FigMonkeyRF}A, B) with an average diameter of 92.8 $\pm$ 10 $\mu$m (SEM, n$=$9). 
Using the same strategy as in the mouse retina, a LN model could predict well the responses to repeated sequence of a checkerboard stimulus on cells with a clear modulation of the firing rate (fig. \ref{FigMonkeyRF}C, Pearson correlation 0.74 $\pm$ 0.04, n$=$3), and was close to the performance expected given the reliability of the response (fig. \ref{FigMonkeyRF}D). \\

\subsection*{Acuity estimation of the reactivated retina}

Intuitively, the receptive field size is related to the spatial resolution of the retina. This spatial resolution is connected with the acuity that a blind patient treated with this optogenetic strategy could achieve. However, one would like to make this connection more quantitative. 
Since our model gives a precise account of how ganglion cells respond to visual stimuli, we can construct a full model of how the complete retinal population respond to a stimulus, and simulate the spike trains that the brain will receive from retinal ganglion cells. Thanks to this model, we can estimate the smallest letter size that can still be discriminated by an observer that would have access to these simulated spike trains. This smallest discriminable letter size is a good proxy of the best acuity reachable thanks to this therapeutic strategy. 
To construct a full model of the reactivated retina we assumed that ganglion cells were placed on a squared grid, with a density equal to the density of transfected cells in the experiment. In a previous study, we found that around 40\% of ganglion cells were transfected in the macaque foveal ring (measured from confocal imaging in \cite{Chaffiol2017a}), and the density of ganglion cells in the macaque fovea has been estimated to 51,108 cells/mm$^2$ \citep{Ahmad_2003}. Each neuron was simulated with an identical LN model, with the parameters (STA, non-linearity) chosen to be equal to the average parameters found in the experimental data (see methods). Each neuron in our model was thus identical up to a translation of its receptive field. 

\begin{figure}
\begin{center}
\includegraphics[width=0.8\linewidth]{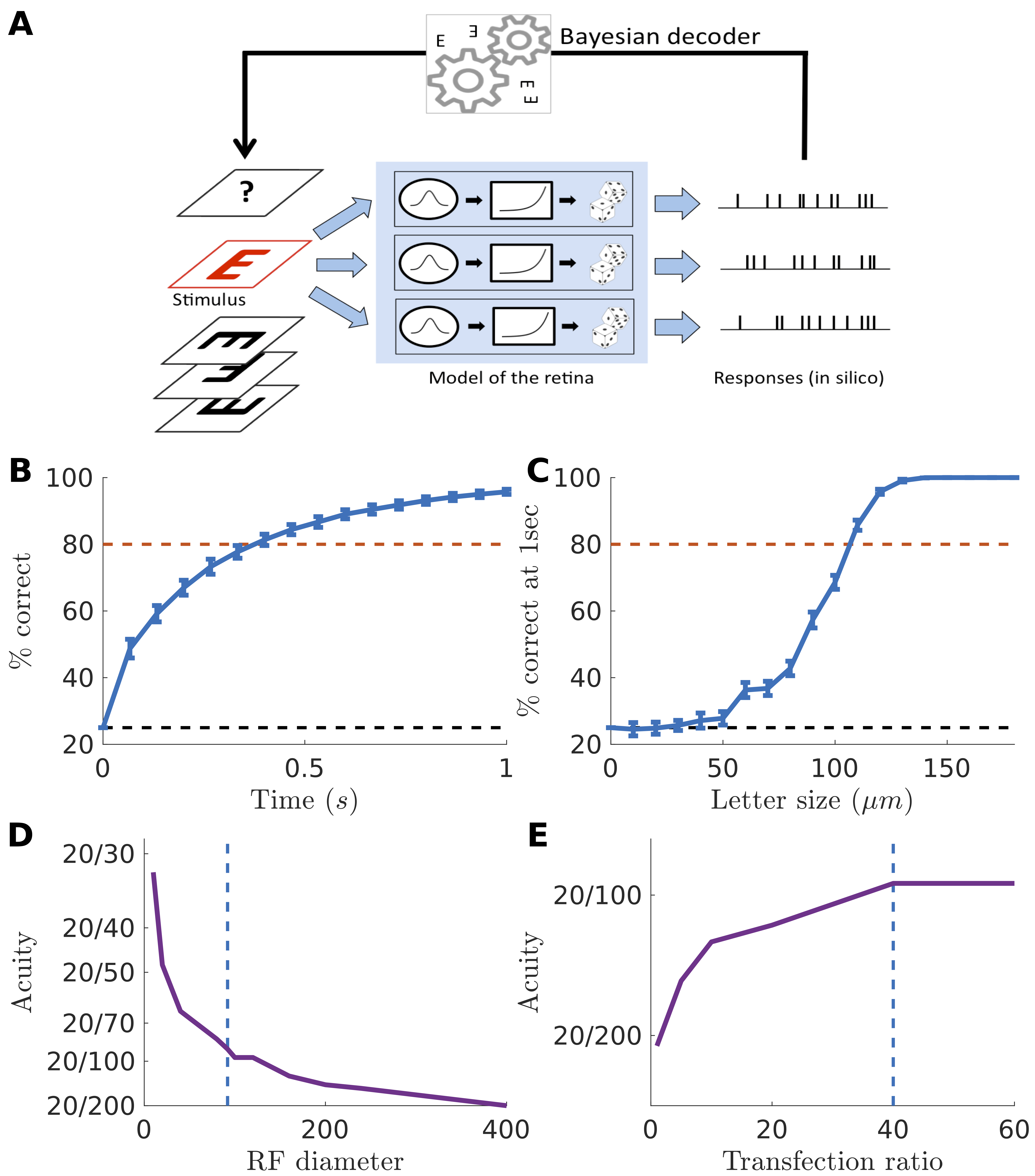}
\end{center}
\caption{ 
\textbf{Simulation of an acuity test \emph{in silico} and estimation of acuity.} (A) We showed the letter E in one of 4 possible orientations and flashed the letter at a new random position every 33 ms to mimic fixational eye movements. We then simulated the entire macaque retinal output as a collection of ganglion cells with receptive fields regularly spanning the visual field, all parameters being fitted to the experiments. To decode which letter was presented from these spike trains, we adopted a maximum likelihood strategy: we assumed perfect knowledge of the model and tested for which stimulus the observed spiking response was the most likely. We then chose the letter with the highest log-likelihood as the prediction of our model. (B) Average success rate over time, for a letter of 120 microns. (C) Percentage of letters decoded correctly after a 1s exposure, against the letter size.The dashed line is the 80\% accuracy limit that we used in our definition of acuity (see results). (D) Predicted acuity as a function of the receptive field size. (E) Predicted acuity as a function of the density of reactivated ganglion cells. 
}
\label{FigPerf}
\end{figure}

We then used our model to simulate the spiking response of the reactivated retina to an acuity test (fig. \ref{FigPerf}A). We chose a classical acuity test used in ophthalmology, the random E test, where the letter `E' is presented in 4 possible directions. The test consisted in presenting randomly a letter to the retina in silico for 1 second, animated by a random jitter mimicking eye movements (see methods). We then predicted which letter was presented from the spike trains using a Bayesian decoder (see methods). By performing a Bayesian inversion of the model, it was possible to estimate which letter was presented. Using this decoder is equivalent to assume that the brain made the best use of the information contained in the spike trains received from the retina (a classical ```ideal observer'' hypothesis). 

We simulated the activity of ganglion cells responding to letters E with different orientations and performed Bayesian decoding at different times following the beginning of the stimulation. As expected the success rate increased with time (\ref{FigPerf}B). This is because the decoder accumulated evidence over time to better discriminate letters, and got better at finding which letter was presented when it had access to longer responses. 
Performance increased with the size of the letter to be decoded (fig.\ref{FigPerf}C). 
To be realistic and consistent with real in-situation acuity tests, we then defined the acuity score as the smallest letter size for which the success rate was above 80\% within a time exposure of 1 second. We found that the smallest discriminable letter size was 110 microns. Assuming that discriminating a letter of 25 $\mu$m gives a 20$/$20 acuity, these 110 microns corresponds to an acuity of 20/88. This is above the threshold of legal blindness ($20/200$).

\section*{Discussion}

\subsection*{Localized receptive fields}
Previous studies using optogenetic proteins expressed in ganglion cells mostly measured responses to full field flashes  \citep{Bi2006,Lin2008,Sengupta2016,Caporale2011,Berry2017}. A few studies where light sensitivity was restored at the bipolar or photoreceptor stage used spot of increasing sizes to determine size selectivity \cite{Lagali2008,Busskamp2010}. However, systematic estimation of receptive fields has not been performed previously when ganglion cells are reactivated. Direct measurements of acuity using behavioural tests have only been performed on mice \citep{Berry2017}. Measuring the acuity with behavioral experiments on non-human primates is made difficult by the fact that the light stimulation necessary to activate the transfected cells will also activate the photoreceptors, and the effect of photoreceptor versus optogenetic activation cannot be easily separated. 
 \\
A possible limitation of vision restoration with optogenetic reactivation of ganglion cells is that the whole axon of the neuron could become light sensitive, thus creating an unnaturally elongated receptive field. We have measured the receptive fields of retinal ganglion cells that were made light sensitive using an optogenetic protein.  
We found that many cells showed a receptive field with a circular shape and a diameter smaller than in normal retinas. The limited size and circular shape of our measured receptive fields demonstrates that the reactivated ganglion cells are mostly sensitive to the stimulation of their soma and proximal dendritic field, and not to the stimulation of their axon: otherwise we would have measured large and elongated receptive fields encompassing the axon image.

\subsection*{Limits of the ideal observer analysis}
To quantify the spatial resolution of the reactivated retina, we built a model of the information transmitted by reactivated ganglion cells to the brain. This model was realistic and properly fitted to the data, and allowed quantitative predictions of the best acuity we can expect in a treated patient. Our model predicted that a patient should be able to discriminate letters of size corresponding to a visual acuity of 20/88. The main assumption in this approach is that the brain can make the best use of the information transmitted by the retina. Even for late blind patients who have already experienced visual stimulation, the stimulation received from the reactivated retina will have a novel structure that needs to be learned. For example, former OFF ganglion cells now respond to light onset, and need to be processed like ON ganglion cells. Learning to use this novel retinal code will require a reorganization of visual cortices, where ON and OFF subregions have a distinct topographic arrangement \citep{Kremkow_2016,LeeK_2016}. A possible consequence is that patients will only use the information coming from ON cells. In our model, this means that only half of the reactivated cells will send usable information, which is equivalent to divide the density of reactivated cells by two. Our prediction is that the impact of this division will be moderate, and acuity will still be above the legal level of blindness (fig. 3E). 
\\
The reactivated ganglion cells form a ring around the photoreceptors of the fovea, and this novel geometry also needs to be learned by the brain. Many studies have shown that a reorganization of the adult visual cortex is possible following a lesion in the retina \citep{Keck_2008, Keck_2013}. Additionally, a promising strategy is to pre-process the visual input before sending a stimulation pattern to the ganglion cells. This pre-processing can be optimized to help the brain make the best use of the information transmitted by the retina. Both pre-processing of the visual input and brain plasticity could help to achieve a perceptual performance close to the optimal spatial resolution estimated here, but only direct tests on patients will determine how close to optimal they can be. \\

\subsection*{Predicting efficacy of future treatments} 

An advantage of our modeling strategy is that it allows varying the different parameters to predict how they will influence the acuity.  Here we have tested the impact of two relevant parameters in this model: the density of cells and the size of the receptive field. These two parameters will change depending on the therapeutic strategy adopted. 

The size of the receptive field could change depending on how protein expression is engineered. \cite{Baker2016} and \cite{Shemesh2017a} have shown novel engineered opsins can restrict their expression in the somas. This could reduce the receptive field diameter and dramatically enhance acuity. Thanks to our modeling approach, we could estimate how acuity should change as a function of the RF diameter (fig. 3D). Very high acuity levels, above 10$/$20, could be reached if the receptive field diameter gets close to soma size (i.e. around 10 microns). However, if this comes at the cost of a lower expression level, resulting for example in a lower transfection ratio, this performance could be mitigated. 

The density of cells will vary if the ratio of transfected cells is varied. This could happen when the AAV dose is changed, if a different capsid, promoter or optogenetic protein is used. We have estimated how the acuity will change when the transfection ratio is changed (fig. 4E). Surprisingly, the predicted acuity is only marginally affected by the transfection ratio, if all the other parameters of the model are kept equal. This is due to the high density of ganglion cells around the fovea: even with a 10 \% ratio a lot of cells will be activated, and this will be enough to transmit information. However, below this ratio performance decreased significantly. 

Our study thus emphasizes that several factors play a role in the expected acuity. 
Our approach integrates these different factors into a coherent framework and has thus the potential to help refining future optogenetic strategies by predicting the expected acuity in various conditions. 

In conclusion, our study shows that the optogenetically activated ganglion cells have a small receptive field. This localized processing enables them to deliver a high resolution picture of the visual scene to the brain. 
Our results suggest that optogenetic therapy based on in vivo AAV injection targeting ganglion cells should give an acuity above the limit of legal blindness. This would be a significant improvement compared to current strategies based on retinal implants, where acuity has remained below that level so far (\cite{Humayun_2012,Stingl_2013}; but see \cite{Lorach_2013} for an alternative approach). Our approach can also be used to predict the impact of refined optogenetic strategies on the predicted acuity, and should be useful for further improvements of this therapeutical strategy. 

\section*{Material and methods}
Unless stated otherwise, all error bars in figures and text are standard deviations over the samples. SEM stands for standard error of the mean and SD for standard deviation.

\subsection*{AAV production and injection}
Details of the gene delivery and optogenetic protein expression in mice has been detailed elsewhere \citep{Sengupta2016,Chaffiol2017a}. Briefly, we targeted retinal ganglion cells (RGCs) of blind rd1 mice (4-5 weeks old) with an AAV2 encoding ReaChR-mCitrine under a pan-neuronal hSyn promoter via intravitreal injections. Four weeks post-injection, we observed strong retinal expression of mCitrine in rd1 mice as revealed by in vivo fundus imaging. 
For macaques, we targeted retinal ganglion cells (RGCs) with an AAV2 encoding a human codon optimized CatCh under a strong, RGC-specific promoter \citep{Chaffiol2017a}. Retinas were harvested three months after injection of the virus in the adult macaque retina.\\
 
\subsection*{Multielectrode array recordings}
Recordings were made using a multielectrode array (MEA) comprised of 252 extracellular electrodes spaced at 100 $\mu$m on a square grid (Multi-Channel Systems, Reutlingen, Germany). Once a piece of retina had been isolated, it was placed ganglion cell side down onto the array. A perforated dialysis membrane was used to hold the retina in place on the array. The array was superfused with Ames solution (3 ml/minute, gassed with 95\% O2-5\% CO2) and maintained at 34$^{\circ}$C. Raw RGC activity was amplified and sampled at 20kHz. Resulting data was stored and filtered with a 200 Hz high pass filter for offline analysis. The recordings were sorted using custom spike sorting software developed specifically for these arrays \citep{Marre_2015,Yger_2016}. 

\subsection*{Visual stimulation and receptive field estimation}

The stimulus was displayed using a Digital Mirror Device and focused on the photoreceptor plane using standard optics \citep{Deny2017}.

The receptive field (RF) of a retinal ganglion cell (RGC) is the particular region of the visual field in which a stimulus will trigger the firing of the cell. Here we characterized the spatial and temporal components of the RFs by estimating the spike-triggered average (STA) from a white noise checkerboard stimulus \citep{Chichilnisky_2001}. 
The stimulus was a flickering black-and-white checkerboard where the intensity of each checker was drawn at random from a binary distribution at every stimulus frame. 
The size of the checks was $67\mu m$ for macaque and $50 \mu m$ for mouse, and frames were updated at respectively $30Hz$ and $40Hz$.
Computing the STA consists in selecting and averaging the frames in a 200ms time window preceding each spike, to form a 3 dimensional description of the receptive field (2 dimensions are space, 1 dimension is time). The spatial RF is defined as the temporal slice of the STA that contains the maximal value of the whole STA. The temporal RF is defined as the temporal evolution of the check of the STA with the maximal average value. To estimate the diameter of the receptive field we fitted a two-dimensional Gaussian to the measured spatial receptive field. \\

Across all spike-sorted cells, we selected the cells that had a salient receptive field: $9/140$ cells for the macaque, and $30/63$ for the mouse, passed this test The small ratio in the macaque is due to the fact that many recorded neurons were outside of the foveal ring, and therefore not transfected efficiently by the AAV \citep{Chaffiol2017a}. 
We then asked if cell responses were robust across repetitions of the same visual stimulus. 
To test for this, we divided stimulus repetitions in two halves and compute the mean response in time (PSTH) for each of them. 
Cells that showed a Pearson correlation between the two PSTHs larger than $0.5$ passed the test: $3/9$ for the macaque and $24/30$ for the mouse. 

\subsection*{Linear Non-linear model}

We fitted the responses of the ganglion cells with a Linear-Non-linear model \citep{Chichilnisky_2001}. 
In this model the stimulus is first convolved with the receptive field of the cell. Then the result goes through a non-linearity to predict the firing rate over time. The non-linearity relates the amount of light in the receptive field to the firing rate of the cell. \\

To test the model, we repeated 50 times another sequence of the checkerboard stimulus. We computed the firing rate predicted by our model and compared it to the experimental firing rate of the cell averaged across the 50 trials. To quantify the ability of our model to predict the cell response, we calculated the Pearson correlation (r) between the predicted ($f_{pred}$) and experimental ($f_{exp}$) firing rate. To measure the response reliability of a neuron, we split the repetitions in two halves and computed a PSTH on each half, and we then computed the Pearson correlation coefficient between these two PSTHs.\\

In the population model, the spatio-temporal RF shape was the same for all the neurons, only the center of the cells differed. 
To estimate this averaged RF, we first computed the position of each recorded cell, then averaged the STAs of the re-centered cells.
Then we decomposed this average RF into a spatial and a temporal component.
The first was a two-dimensional, symmetric, Gaussian, whereas the latter was the RF temporal modulation at the center.
For the non-linearity, we parametrized it as a soft-plus function with three parameters \cite{Deny2017} that we fitted with maximum log-likelihood from the response to unrepeated checkerboard data.
We then took the uniform RF and we averaged the non-linearity across the cells to simulate an entire population. 

\subsection*{Acuity test in silico}

To simulate the test, we picked one out of the 4 possible orientations and flashed the letter E at a new random position every 67 ms (corresponding to the decay time constant of the temporal RF). The purpose of this random renewal of the position is to mimic the fixational eye movements of the patient, which would displace the letter over the retinal surface (see also \cite{Burak_2010}). The letter was white on a black background (100\% contrast). \\
We then simulated the entire retinal output as a collection of ganglion cells with receptive fields regularly spanning the visual field. We randomly picked a subset of cells that were the ones supposed to express the optogenetic proteins - the other ones were supposed to send no information about the stimulus and therefore removed. Each time a stimulus was presented, it was convolved with the receptive field of each cell and the result went through the non-linear function to predict the firing rate for each cell. 
We assumed that ganglion cells emitted spikes according to a Poisson process, as in a classical Linear-Non-linear Poisson model, previously used in the retina \citep{Pillow_2008}. \\

We then decoded which letter was presented from these spike trains. For this we adopted a maximum \textit{a posteriori} strategy (with flat prior): we assumed perfect knowledge of the model and tested for which stimulus the observed spiking response was the most likely. First we computed the firing rates of the whole population of cells in response to every possible position and every letter f(cell, position, letter). Then, assuming a Poisson distribution of firing rates, we calculated the log-likelihood of the firing rates observed given any letter \citep{Doya_2007}:

\begin{multline}
\log p(f_{obs}|letter)  = \sum_{position,cell} f_{obs}(cell) \log f(cell, position, letter) \\
- \Delta t \sum_{position,cell} f(cell,position,letter) + c
\end{multline}

where $f_{obs}$ are the firing rates simulated for all cells in response to the letter presented, $\Delta t$ is the time of presentation of the letter in a given position (60ms), and $c$ is a constant. 

We then chose the letter with the highest log-likelihood as the prediction of our model. \\

The decoding was performed at each time step. Over the time course of the presentation, the decoding could benefit from evidence accumulated at previous time steps. The decoding performance thus got better and better over time. We repeated this test 500 times with random letters and averaged the performance over time of our decoder. The performance was defined as the percentage of letters correctly guessed. We defined the acuity as the smallest letter for which the performance was larger than $80$\% after a time exposure of 1 second. 
To estimate the standard deviation in our results, we repeated the whole procedure 30 times.

\section*{Acknowledgments}

This work was supported by ANR TRAJECTORY, by the European Union's Horizon 2020 research and innovation programme under grant agreement No 785219, No. 785907 (Human Brain Project SGA2) and No. 639888, NIH grant U01NS090501, a grant from AVIESAN-UNADEV to OM, a Foundation Fighting Blindness grant to S.P, D.D. and J.A.S., ERC Starting Grant (OPTOGENRET, Grant No 309776) to JD and ERC Starting Grant REGENETHER to D.D., and by the French State program Investissements d'Avenir managed by the Agence Nationale de la Recherche [LIFESENSES: ANR-10-LABX-65].  S.D. was supported by a PhD fellowship ``DIM cerveau et pensee'' from the region Ile-de-France. 

\bibliographystyle{apalike}
\bibliography{BIBLIO,AcuityPaper}

\end{document}